\documentclass[pra,aps,amsmath,amssymb,amsfonts,superscriptaddress,twocolumn]{revtex4-2}
\usepackage{graphicx,color,mathtools,bm,braket,newtxtext,newtxmath}

\providecommand{\abs}[1]{\left\lvert#1\right\rvert}

 \usepackage[T1]{fontenc}
\usepackage{dsfont}

\usepackage[hyphenbreaks]{breakurl}
\usepackage[colorlinks=true,linkcolor=blue,citecolor=blue,urlcolor =blue]{hyperref}

\usepackage{appendix}
\usepackage{amsfonts,amssymb}
\usepackage{physics}
\usepackage{tikz}
\usepackage{float}

\usepackage{amsmath}
\usepackage{bbold}
\usepackage{hyperref}
\hypersetup{
     colorlinks=true,      
    linkcolor=blue,        
    citecolor=blue,        
    filecolor=magenta, 
    urlcolor=black          
}

\usepackage{comment}

\usepackage{amsmath}
\usepackage{bbold}
\usepackage{hyperref}
\hypersetup{
     colorlinks=true,      
    linkcolor=blue,        
    citecolor=blue,        
    filecolor=magenta, 
    urlcolor=black          
}

\usepackage{comment}

\renewcommand{\selectlanguage}[1]{}

\begin{document}

\author{Nicolas Fabre}
\email{nicolas.fabre@telecom-paris.fr}
\affiliation{Telecom Paris, Institut Polytechnique de Paris, 19 Place Marguerite Perey, 91120 Palaiseau, France}
\author{Ulysse Chabaud}
\affiliation{DIENS, \'Ecole Normale Sup\'erieure, PSL University, CNRS, INRIA, 45 rue d'Ulm, Paris 75005, France}
 
\date{\today}


\begin{abstract}
The celebrated Hong--Ou--Mandel effect illustrates the richness of two-photon interferometry.  In this work, we demonstrate that this extends to the realm of time-frequency interferometry. Taking advantage of the mathematical analogy  which can be drawn between the frequency and quadrature degrees of freedom of light when there is a single photon in each auxiliary mode, we consider the equivalent of the Hong--Ou--Mandel effect in the frequency domain. In this setting, the $n$-Fock state becomes equivalent to a single-photon state with a spectral wave function given by the $n^{th}$ Hermite--Gauss function and destructive interference corresponds to vanishing probability of detecting single photons with an order one Hermite--Gauss spectral profile. This compelling analogy motivates us to propose an interferometric strategy that uses a frequency-engineered two-photon state to achieve enhanced phase precision that scales inversely with the number of modes. Finally, we generalise the Gaussian Boson Sampling model to time-frequency degrees of freedom of single photons.
\end{abstract}




\title{Photonic quantum information processing using the frequency continuous-variable of single photons}


\maketitle


\section{Introduction}

\begin{table*}[!htbp]
   \begin{tabular}{|c|c|}
  \hline
  \textbf{Quadratures variables} & \textbf{Time-frequency variables of single photons} \\
  \hline\hline
 $\hat{X}_{a}, \hat{P}_{a}$,  $[\hat{X}_{a},\hat{P}_{a}]=\hat{\mathds{I}}$ & $\hat{\omega}_{a}, \hat{t}_{a}$, $[\hat{\omega}_{a},\hat{t}_{a}]=\hat{\mathds{I}}$ \\
  \hline
 $\hat{X}_{a}\ket{X}_{a}=X\ket{X}_{a}$, $\hat{P}_{a}\ket{P}_{a}=P\ket{P}_{a}$ &  $\hat{\omega}_{a}\ket{\omega}_{a}=\omega\ket{\omega}_{a}$, $\hat{t}_{a}\ket{t}_{a}=t\ket{t}_{a}$ \\
  \hline
  Vacuum state & Single photon with a Gaussian spectrum \\
  \hline
  $N$-Fock state & Single photon with a Hermite--Gauss spectral function of order $N$ \\
  \hline
  $\hat{U}_{BS}\ket{X_{s},X_{i}}_{ab}= \ket{\frac{X_{s}+X_{i}}{\sqrt{2}},\frac{X_{s}-X_{i}}{\sqrt{2}}}_{ab}$ & $\hat{U}\ket{\omega_{s},\omega_{i}}_{ab}= \ket{\frac{\omega_{s}+\omega_{i}}{\sqrt{2}},\frac{\omega_{s}-\omega_{i}}{\sqrt{2}}}_{ab}$ \\
  \hline
\end{tabular}
\caption{Mathematical analogy between the quadratures and the time-frequency degrees of freedom of single photons. The analogy holds as soon as the auxiliary modes $a,b$ are populated by one single photon.}
    \label{inte}
\end{table*}

Photonic quantum information processing is based on playing with fundamental bosonic properties of quantum states of light. 
The celebrated Hong--Ou--Mandel (HOM) bunching effect \cite{hong1987measurement} provides one of the most basic but also striking examples of the nonclassical behaviour of photons.
Boson Sampling \cite{bs}, a computational generalisation of the HOM effect, has been introduced as a near-term candidate for harnessing the computational power of photonics \cite{wang_boson_2019,bentivegna_experimental_2015}. 
As such, its variant Gaussian Boson Sampling (GBS), defined in \cite{gbs}, led to the first attempts of demonstration of experimental quantum computational advantage using photonic platforms \cite{ZHONG2019511, madsen_quantum_2022,PhysRevLett.131.150601}. 

The theoretical protocol for GBS involves the preparation of a multimode Gaussian quantum state and its measurement by a photon number-resolving detector in each mode. The probability of obtaining a given photon-number outcome is related to a combinatorial function known as the hafnian, a quantity that is $\#$P-hard to compute \cite{Valiantpermanent}. 
This approach directly utilizes entanglement, interference phenomena, and non-Gaussian operations to gain a potential quantum advantage over classical algorithms. The hafnian is associated with various mathematical problems in graph theory \cite{PhysRevLett.130.190601,PRXQuantum.5.020341}, including graph isomorphism \cite{isomorphismgbs}, the clique problem, perfect matching of undirected graphs, and finding dense subgraphs \cite{densesubgraphgbs,PhysRevX.12.031045}. As a result, GBS has potential applications in quantum chemistry \cite{chemistry,vibronicgbs} and quantum optimization \cite{optimizationgbs}, although the precise quantum advantage of GBS for specific computational tasks beyond sampling remains uncertain.

Many quantum optics platforms focus on degrees of freedom of quantum states of light such as polarization, spatial modes or time-bins, including the recent GBS experiments \cite{ZHONG2019511, madsen_quantum_2022}. However, quantum information can be also encoded using the time and frequency degrees of freedom of single photons, that are intrinsically continuous variables \cite{PhysRevA.42.4102,fabre_generation_2020}, but can be discretized into bins or modes. Recent advances have considerably improved the manipulation of the frequency (or spectral) degree of freedom of single photons: techniques that use pulse shapers and/or electro-optic modulators \cite{lukens_frequency-encoded_2017,PhysRevA.107.062610,kurzyna_variable_2022,PhysRevLett.130.240801,imany_high-dimensional_2019,kues_quantum_2019}, or quadrature phase‑shift keying modulators \cite{chen_single-photon_2021}, allow for performing single photon operations, as well as two-photon operations \cite{lu_controlled-not_2019,Le_Jeannic_2022}; other techniques for manipulating the time-frequency degree of freedom of photon pairs modify the temperature of a non-linear crystal at the generation stage \cite{chen_hong-ou-mandel_2019,jin_quantum_2018}, by engineering the spatial distribution of the pump as in integrated circuits \cite{francesconi_engineering_2019,maltese_generation_2020,francesconi_anyonic_2021}, in atomic cloud, or by engineering the spatial structure of the non-linear crystal \cite{dosseva_shaping_2016,jin_generation_2016,luo_counter-propagating_2020,gao_manipulating_2022}. 

In \cite{fabre_time-frequency_2022}, a formalism to establish the mathematical correspondence between the quadrature and the frequency degrees of freedom considered as a continuous variable was established, based on the property that  each  auxiliary mode (other than the frequency) must be occupied by one single photon. This formalism opens new avenues for quantum information processing, enabling universal quantum computations by utilizing the frequency degree of freedom of single photons as a continuous-variable encoding of quantum information. Furthermore, it facilitates an understanding of the spectral (or temporal) width as a source of quantum noise in computing, as explored in earlier work \cite{fabre_generation_2020,fabre_teleportation-based_2023}, and in recent quantum metrology protocols \cite{PhysRevLett.131.030801}.

In this paper, based on this mathematical analogy between the quadrature and frequency continuous variables, we consider the equivalent of the HOM effect in the frequency domain of single photons. Following this analogy, $N$ excitations of the electromagnetic field correspond to single-photon states with a Hermite--Gauss frequency spectrum of order $N$. The beam-splitter in the original HOM effect is replaced by a non-linear frequency beam-splitter (NLFBS) which entangles the frequency of two single-photon states occupying two auxiliary modes (already mentioned in \cite{fabre_time-frequency_2022,doi:10.1080/09500340.2022.2073613,fabre_generation_2020}), and performs a $\pi/4$ rotation of the joint spectral amplitude of two single photons. The final output is an intriguing frequency interference phenomena that suppress single photons in one frequency mode, similar to the destructive interference effect that leads to bunching in the HOM experiment. The proposed experiment in this paper can be seen as the continuous-variable counterpart of the one described in \cite{imany_frequency-domain_2018} using linear optics or with quantum frequency translation \cite{mcguinness_theory_2011}.

We then provide a first application of the frequency-based HOM effect in quantum metrology. Following the same analogy, we then define a quantum state composed of two-photon states  with a spectrum that is the Hermite--Gauss function of order $N$ (see \cite{PhysRevA.65.052104,dowling_quantum_2008} and develop the strategy developed adapted from \cite{holland_interferometric_1993} for reaching a precision that scales as the number of mode  for phase estimation $\Delta \phi \sim 1/N$, similar mathematically to a Heisenberg-like limit.  This quantum-inspired strategy was developed using the orbital angular momentum of one single photon \cite{dambrosio_photonic_2013}. Performing spectral engineering of photon pair is relevant for two-photon metrological scenario and adapted for probing fragile materials such as biological ones \cite{doi:10.1126/sciadv.aap9416,fabre_parameter_2021}, and improving signal-to-noise ratio.

Finally, our mathematical correspondence finds application in quantum computing through the novel concept of GBS generalized in the continuous time-frequency variables of single photons. In this protocol, we start from $N$-single photons that are present in $N$ distinct auxiliary modes (such as spatial modes) with a Gaussian spectrum (that are mathematically equivalent to squeezed states), that enter into an interferometer performing time-frequency Gaussian operations. Frequency  entanglement operations are performed with \textit{non-linear} operations through light-matter interaction. The output state is a time-frequency Gaussian state whose chronocyclic Wigner distribution \cite{dorrer_concepts_2005,brecht_characterizing_2013,fabre_generation_2020} is Gaussian. The final step involves single photon mode-resolved detection, which is projection onto a spectral Hermite--Gauss basis, coupled with measurements employing single-photon detectors, thereby simulating the equivalent of photon number-resolving detection. A given configuration is given here as the simultaneous measurement of $N$ single-photon states with Hermite--Gauss mode spectrum $n_1,...n_N$. In time-frequency GBS, a configuration involves the measurement of specific single-photon states, each associated with its respective Hermite--Gauss mode spectrum, in contrast to Boson sampling where configurations are defined by the detection of single photon states, irrespective of their degrees of freedom structure. The output distribution is the hafnian of a symmetric matrix, that is a $\#$P-hard problem and we therefore obtain a quantum computing model that samples from the distribution of photons in a \textit{non-linear} optical network. While we outline prospects for implementing our proposal experimentally, our main focus is rather to highlight that different types of quantum resources and degrees of freedom can lead to the same statistics and information processing capabilities.

The rest of the paper is organized as follows. In Sec.~\ref{FBQIP}, we develop the essential tools for understanding the formalism describing the frequency degree of freedom of single photons. In Sec.~\ref{HOMsec}, we discuss the continuous-variable frequency HOM effect by using the mathematical analogy between the quadrature and the frequency degree of freedom. Following this analogy in Sec.~\ref{NOONsec}, we then develop a quantum metrology protocol relying on a two-photon probe whose frequency spectrum is engineered so that to achieve Heisenberg scaling for phase estimation. Finally in Sec.~\ref{GBS}, we translate the GBS model to the frequency degree of freedom of single photons.


\section{Frequency-based quantum information processing}
\label{FBQIP}

In this section, we recall the formalism of a single photon with a continuous-variable frequency degree of freedom and refer the reader to Appendix~\ref{details} for more details. 

A single photon with a frequency $\omega$ in the spatial mode (or any other auxiliary mode) $a$ is denoted $\hat a^\dag(\omega)\ket\Omega_a$, where $\ket\Omega_a$ is the vacuum state. The Fourier transform of a creation operator in the frequency domain correspond to a creation operator in the temporal domain. We introduce the frequency and time operators as:
\begin{align}
    \hat{\omega}_{a}&=\int_{\mathds{R}} d\omega\,\omega\,\hat a^\dag(\omega)\hat a(\omega),\\
    \hat t_a&=\int_{\mathds{R}} dt\,t\,\hat a^\dag(t)\hat a(t).
\end{align}
Although frequencies are nonnegative, the integration is over $\mathds{R}$ without loss of generality: in practice, wave functions are localized far away from the origin. If the auxiliary mode is populated by a single photon, the frequency and time operator verify the Heisenberg--Weyl canonical commutation relation \cite{fabre_time-frequency_2022}:
\begin{equation}\label{heisenberg}
    [\hat{\omega}_{a},\hat{t}_{a}]=\hat{\mathds{I}}.
\end{equation}
Therefore, we must have in each spatial mode only one single photon for the mathematical analogy between the quadrature and frequency variables to hold. Equipped with this mathematical analogy, summarised in Table~\ref{inte}, a single photon with Gaussian frequency spectrum becomes mathematically analogous to the vacuum state in the quadrature domain:
\begin{equation}\label{zerofrequency}
    \ket0_{a}\equiv\frac{1}{(\pi\sigma^{2})^{1/4}} \int_{\mathds{R}} d\omega e^{-\omega^2/2\sigma^2} \ket\omega_{a},
\end{equation}
where $\sigma$ is the frequency width of the Gaussian spectrum. A single photon spectrum is centered around a certain frequency that we will omit, which depends on the physical system producing such a single photon. Such a state would be actually the equivalent of a coherent state in the quadrature domain, but as we will set the central frequency to zero, the state becomes equivalent to the vacuum. Then, the $n^{th}$ Fock state is in this encoding a single photon with a Hermite--Gauss frequency spectrum corresponding to the wave function \cite{fabre:tel-03191301}:
\begin{equation}\label{Focknew}
    \ket{n}_{a}\equiv\frac{1}{(\pi\sigma^{2})^{1/4}}\int_{\mathds{R}} d\omega \frac{1}{\sqrt{2^{n}n!}} H_{n}(\omega/\sigma)  e^{-\omega^{2}/2\sigma^{2}} \ket{\omega}_{a},
\end{equation}
where for instance $H_0(x)=1$, $H_{1}(x)=2x$ and $H_{2}(x)=4x^{2}-2$. We can check that $\bra{n}\ket{m}=\delta_{nm}$. Projecting the state $\ket{0}$ into $\bra{n}$ reads as applying a Hermite--Gauss filter $H_{n}(\omega/\sigma)e^{-\omega^{2}/2\sigma^{2}}/\sqrt{2^{n}n!}$, that could be achieved with spatial light modulators and gratings (see for instance \cite{goel_inverse_2024,yan_generation_2022}), by first mapping the frequency to the spatial degree of freedom.\\

The projector defined as $\hat{\Pi}(n)=\ket{n}\bra{n}$ is the analogous of photon-number resolving detectors and will be called mode-resolved single photon detection. The probability distribution obtained with the Fock-like basis is therefore: $\langle \hat{\Pi}(n) \rangle=\abs{\bra{n}\ket{\psi}}^{2}$. This could be achieved by selecting the $n$th Hermite--Gauss mode. Note that a similar mapping from Fock states to Hermite--Gauss modes of the \textit{spatial} degree of freedom of single photons was previously proposed \cite{abouraddy_violation_2007}, together with proposal for the violation of a Bell inequality.\\

\begin{figure}\label{fbss}
    \begin{center}
        \includegraphics[width=0.5\textwidth]{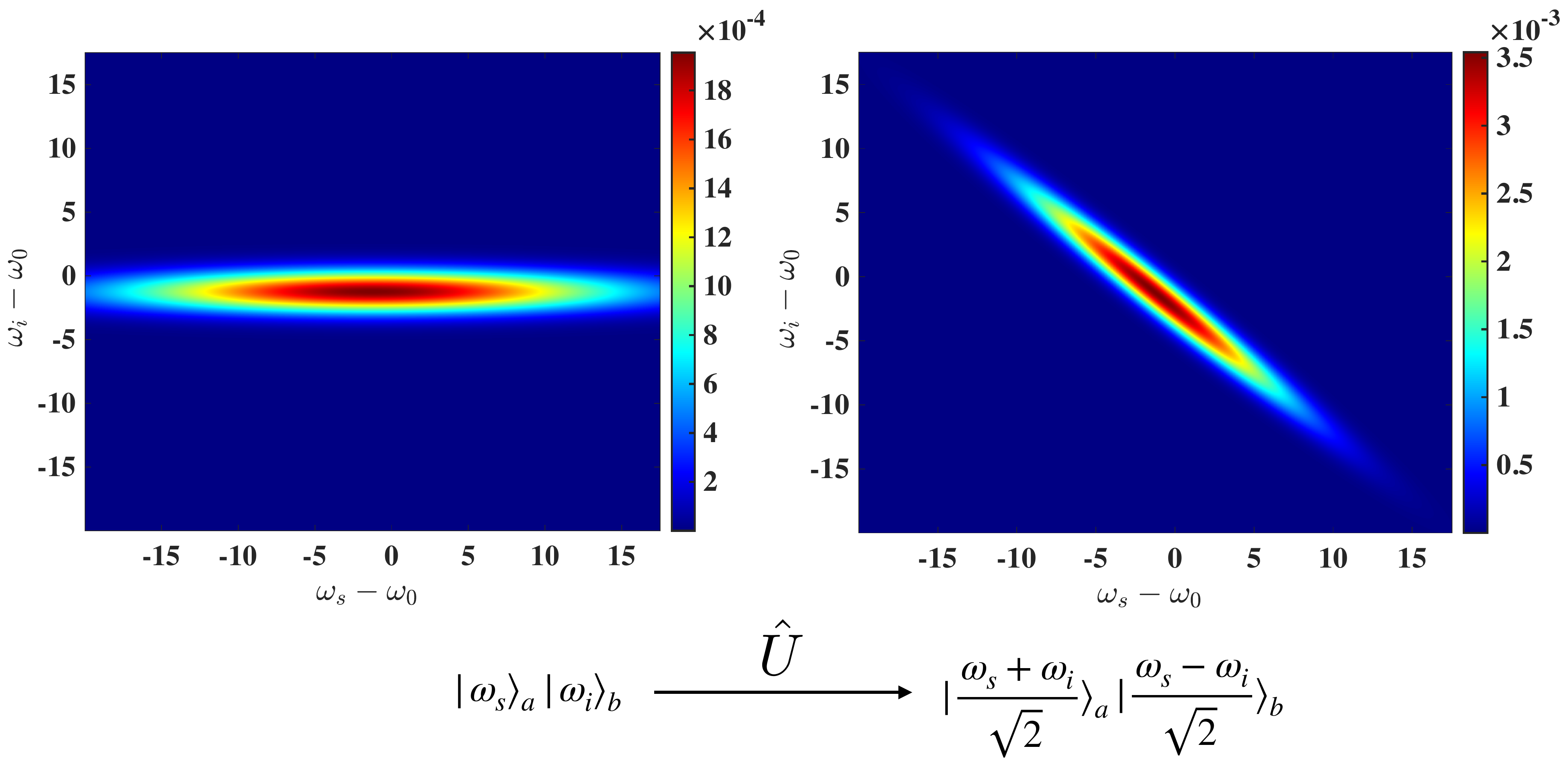}
        \caption{\label{fbsss}Two initially separable single photons have Gaussian frequency spectra with different widths, and their joint spectral intensity is represented on the left. After the action of a non-linear frequency beam-splitter, the photon pair is now frequency entangled, and the non-linear frequency beam-splitter acts as a $\pi/4$ rotation of the spectrum. $\omega_{0}$ is a given central frequency. The equivalent process in the quadrature domain is to start with two single-mode squeezed states which become an entangled two-mode squeezed state after a standard beam-splitter operation.}
    \end{center}
\end{figure}

In the frequency domain, the equivalent of the beam-splitter is the NLFBS \cite{doi:10.1080/09500340.2022.2073613,fabre_generation_2020,fabre_time-frequency_2022} which performs the operation (see Appendix \ref{appendix:FBS} for more details):
\begin{equation}\label{FBS}
    \hat{U}\ket{\omega_{s},\omega_{i}}_{ab}= \ket{\frac{\omega_{s}+\omega_{i}}{\sqrt{2}},\frac{\omega_{s}-\omega_{i}}{\sqrt{2}}}_{ab}.
\end{equation}
Such an operation performs a $\pi/4$ rotation of the joint spectral amplitude of two single photons, leading them to frequency entanglement if the initial joint spectral amplitude was separable (see Fig.~\ref{fbsss}). Besides, each input and output modes ($a,b$) are occupied by one and only one single photon. Akin to the beam-splitter in the quadrature domain (see Table~\ref{inte}), this is a Gaussian operation for the frequency continuous variable \cite{fabre:tel-03191301,Descamps_2023}. The NLFBS has not yet been achieved experimentally, as this is an operation whose Hamiltonian is quartic in bosonic operators. Nevertheless, an operation that leads to the same frequency entangled output has been achieved experimentally by using a quantum dot embedded into a waveguide where the efficiency of one process is of 99\% \cite{Le_Jeannic_2022} (see also \cite{PRXQuantum.4.030326} for further theoretical study). In these studies, instead of performing the full rotation of the joint spectral amplitude of the two photons, the initial photon pairs interact with the quantum dot, that leads to a final joint spectral amplitude that is frequency-carved along the antidiagonal (as in Fig.~\ref{fbsss}), with the output photon pair being entangled in frequency. 

Note that \cite{jones_photon_2006,imany_frequency-domain_2018} define a different beam-splitter in frequency, where the two spatial modes of the standard beam-splitter are now two frequency modes. We also emphasize that the non-linear frequency beam-splitter considered in this work is not a frequency-dependent beam-splitter \cite{makarov_quantum_2021,makarov_theory_2022}, since such a device acts on the quadrature as a standard beam-splitter but with frequency-dependent reflection coefficient.

Another time-frequency gate that will be important in this paper is the fractional Fourier transform  \cite{fabre:tel-03191301} that can be cast as:
\begin{equation}\label{fractional}
    \hat{F}(\phi)=\text{exp}(i\phi(\hat{\omega}^{2}+\hat{t}^{2}-\hat{\mathds{I}})),
\end{equation}
and which has been demonstrated experimentally in \cite{PhysRevLett.130.240801}. Note that we have employed dimensionless units for the frequency and time operators, as it was done in the transversal position-momentum of single photons \cite{PhysRevA.83.052325}. The fractional Fourier transform in the frequency domain is equivalent to the phase-shift operation $\smash{e^{i\phi\hat{b}^{\dagger}\hat b}}$ in the quadrature domain. Therefore, the action of the fractional Fourier transform on a single photon with a $n$th-order Hermite-Gauss function is:
\begin{equation}
  \hat{F}(\phi)\ket{n}_{a}=e^{in\phi}\ket{n}_{a}.
\end{equation}

Note that defining dimensionless position and momentum variables relies on the curvatures of position and momentum during free space propagation and the application of lenses \cite{PhysRevA.83.052325}. In the case of time-frequency encoding, analogous operations can be performed using frequency chirp (achieved via spatial light modulators or optical fibers) and temporal chirp (using time lenses). By considering the curvature parameters associated with the frequency chirp and temporal chirp, we can similarly define dimensionless time and frequency operators.


\section{HOM effect in the frequency domain of single photons}
\label{HOMsec}

Let us assume that we start with two separable single photon in spatial paths $a,b$ that possess the same Hermite--Gauss spectrum $\ket\psi=\ket{1,1}_{a,b}:=\ket1_a\otimes\ket1_b$ (see Eq.~\ (\ref{Focknew})), and let us proceed to the equivalent of the HOM experiment: the state $\ket\psi$ is sent into a non-linear frequency beam-splitter (see Eq.\ (\ref{FBS})). After such an operation, we remain in the single-photon subspace for each output spatial mode. The output wave function after the non-linear frequency beam-splitter is given by:
\begin{multline}\label{HOM}
    \hat{U}\ket{\psi}=\frac{1}{(\pi\sigma^{2})^{1/2}}\iint d\omega d\omega'  \frac{1}{2} H_{1}\left(\frac{\omega}{\sigma}\right) H_{1}\left(\frac{\omega'}{\sigma}\right)  \\ \cross e^{-\omega^{2}/2\sigma^{2}}e^{-\omega'^{2}/2\sigma^{2}}\ket{\frac{\omega+\omega'}{\sqrt{2}},\frac{\omega-\omega'}{\sqrt{2}}}_{ab}.
\end{multline}
We perform a change of variable to obtain:
\begin{multline}
    \hat{U}\ket{\psi}=\frac{1}{(\pi\sigma^{2})^{1/2}}\iint d\omega d\omega'  \frac{1}{2} H_{1}\left(\frac{\omega+\omega'}{\sqrt{2}\sigma}\right) H_{1}\left(\frac{\omega-\omega'}{\sqrt{2}\sigma}\right)   \\
 \cross   e^{-(\omega+\omega')^{2}/4\sigma^{2}} e^{-(\omega-\omega')^{2}/4\sigma^{2}}   \ket{\omega,\omega'}_{ab}.
\end{multline}
In the standard HOM interference, the output wave function written with quadrature variables is not often employed, however, in the particle-number representation the output state is given by
\begin{multline}
    \frac1{\sqrt2}(\ket{2,0}_{ab}-\ket{0,2}_{ab}+\ket{1,1}_{ab}-\ket{1,1}_{ab})\\
    =\frac1{\sqrt2}(\ket{2,0}_{ab}-\ket{0,2}_{ab}),\quad\quad\quad\quad
\end{multline}
which clarifies the origin of the destructive interference between coincidence events.

\begin{figure*}
    \begin{center}
        \includegraphics[width=\textwidth]{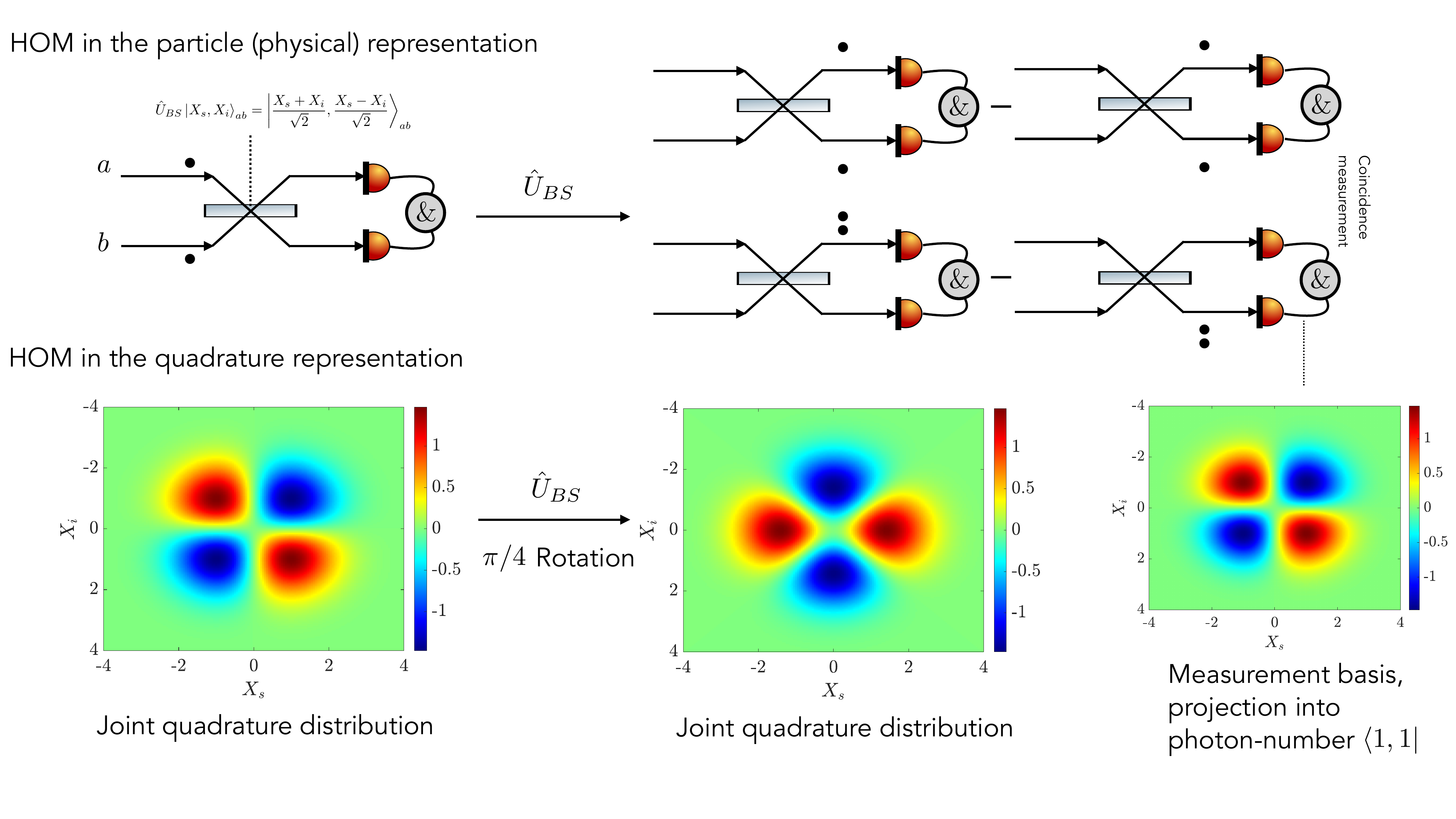}
        \caption{\label{usualHOM}Standard HOM experiment, analysed both in the particle representation and by representing the joint quadrature distribution at each step. Two indistinguishable single photons are combined on a balanced beam-splitter (white rectangle) that is modeled by the unitary operation $\hat{U}_{\text{BS}}$. Coincidence measurement is performed in both arms of the interferometer.  In the quadrature representation, the state that was initially separable becomes entangled after the beam-splitter. As the input and the output are orthogonal - it corresponds to a rotation of the joint quadrature distribution -, thus the corresponding measured probability is zero.}
    \end{center}
\end{figure*}

\begin{figure*}
    \begin{center}
        \includegraphics[width=\textwidth]{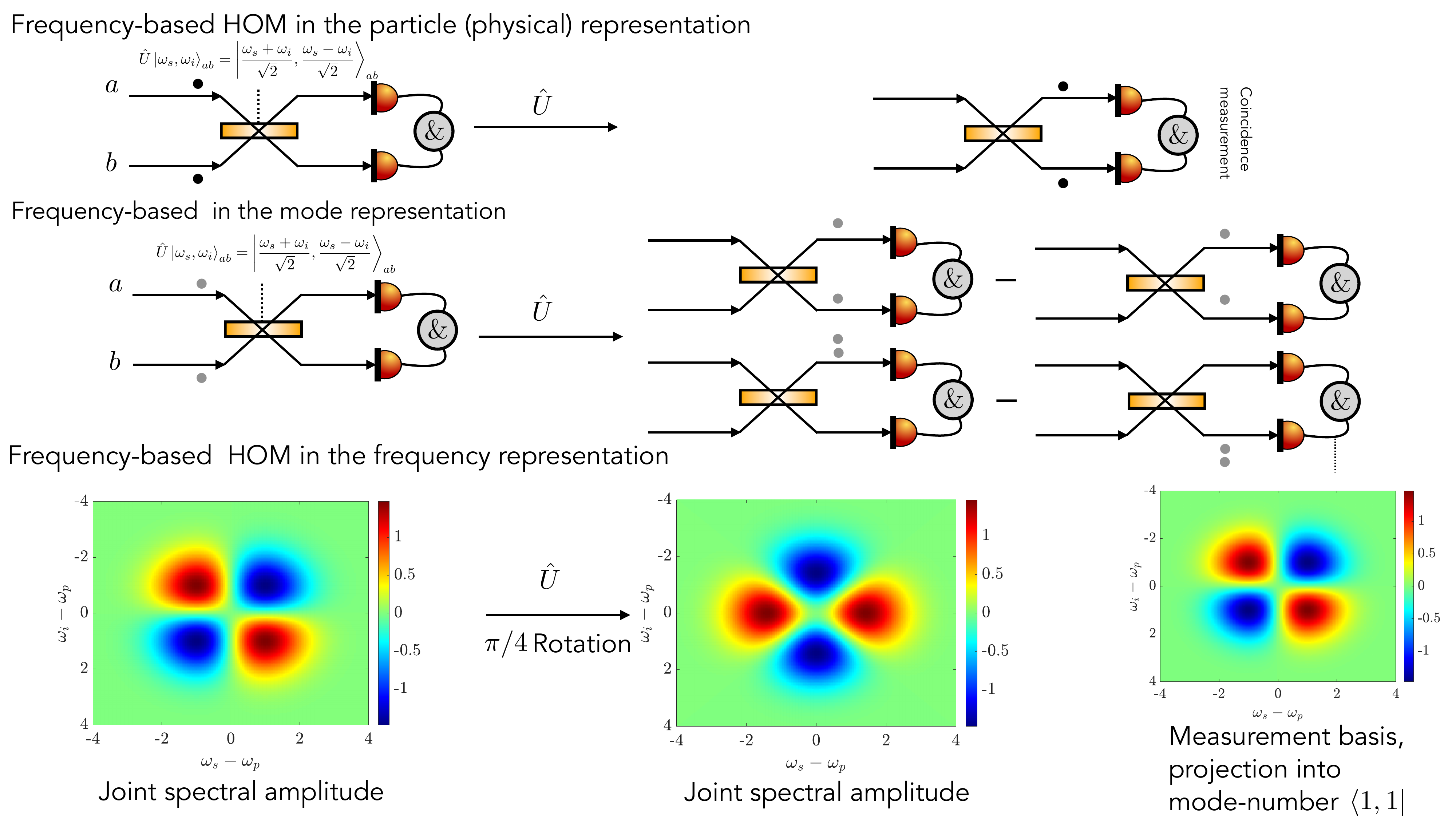}
        \caption{\label{newHOM}Frequency-based HOM experiment using the spectral degree of freedom of single photons. We start from two separable single photons with a Hermite--Gauss spectrum or order one, that become frequency entangled after the non-linear frequency beam-splitter operation $\hat{U}$ (yellow rectangle). We obtain an equivalent destructive interference which does not affect the total particle number in each spatial mode $a,b$ (that stays fixed during the non-linear frequency beam-splitter operation) but the spectral degree of freedom of each single photon. Each path ($a,b$) is always occupied by one single photon, as represented in the particle (physical) representation. In the particle (mathematical) or mode representation, black dots are not single photon, but instead represent the Hermite--Gauss mode labeling of the spectral degrees of freedom, from which there is destructive interference between different spectral contributions.}
    \end{center}
\end{figure*}
In comparison, in the frequency domain, we obtain a destructive interference between different frequency modes, i.e., the equivalent of a ``frequency bunching'' effect. Indeed, from Eq.~(\ref{HOM}), we have $\smash{H_{1}(\frac{\omega+\omega'}{\sqrt{2}\sigma}) H_{1}(\frac{\omega-\omega'}{\sqrt{2}\sigma})=\frac{\omega^{2}}{\sigma^{2}}-\frac{\omega'^{2}}{\sigma^{2}}}$ which  corresponds directly to the two bunched terms, and we do not get any order one polynomial. Projecting the state into the equivalent of coincidences ${}_{ab}\!\bra{1,1}$, which is a photon number-resolving detection in the quadrature domain, corresponds to applying first a spectral first order Hermite--Gauss function and then applying a non-frequency resolved detection (summing over all frequencies) as described in \cite{LEGERO2006253,fabre_producing_2020}. Therefore,  the projection ${}_{ab}\!\bra{1,1}$ corresponds to mode-resolved single photon detection for only the mode $1$. The probability of detecting the photon pair in coincidence in the frequency Hermite-Gauss modes $1,1$ is then:
\begin{equation}
    \abs{{}_{ab}\!\bra{1,1}\hat{U}\ket{\psi}}^{2}=|{}_{ab}\!\bra{1,1} \frac1{\sqrt2}(\ket{2,0}_{ab}-\ket{0,2}_{ab})|^{2},
\end{equation}
which is zero. Thus, we have obtained a non-trivial destructive frequency interference between two single photons having an order $1$ Hermite--Gauss frequency spectrum, which is the analogue of the HOM effect in the frequency domain.

In Fig.~\ref{usualHOM} and Fig.~\ref{newHOM}, we have represented the continuous-variable quadrature (resp.\ frequency) states before and after the beam-splitter (resp.\ non-linear frequency beam-splitter). The beam-splitter has the effect of performing a rotation of the joint spectrum, leading to an entangled state in the quadrature and the frequency domains, respectively.\\

We note that if the two single photons entering the NLFBS exhibit any degree of distinguishability—such as differences in their frequency widths—the resulting destructive interference will be imperfect, leading to a reduced visibility of the interference dip. This phenomenon is analogous to the well-known Hong–Ou–Mandel (HOM) experiment, where any partial distinguishability between two single photons similarly degrades the interference contrast observed in the coincidence probability.


\section{Phase estimation at the Heisenberg limit using two-photon interferometry}
\label{NOONsec}

In this section, we provide a frequency-engineered two-photon quantum state and a new measurement strategy for measuring a phase parameter whose precision scales linearly with the inverse of the number of modes, which is quantum-inspired from \cite{holland_interferometric_1993}.

Heisenberg scaling refers as the scaling of the variance of the phase as $1/N$, $N$ refers as the the number of photons (either NOON state or squeezed state), therefore beating classical resources that is shot-noise limited with a scaling of $1/\sqrt{N}$. Hereafter, the attainable Heisenberg-like scaling is instead defined as linear with respect to the number of modes $N$ and is independent of the photon number since our scheme employs a fixed two-photon resource. This approach parallels earlier demonstrations using the orbital angular momentum degree of freedom in the single-photon regime (see, e.g., \cite{dambrosio_photonic_2013, Bouchard:17}). Rather than providing a strategy to overcome the shot-noise limit, our focus is on a spectral engineering method that enhances the phase resolution achievable with a two-photon state within a quantum-enhanced protocol. Nonetheless, by extending this technique to larger frequency-entangled photon states, it becomes feasible to beat the shot-noise limit, as illustrated in \cite{dambrosio_photonic_2013, PhysRevLett.131.030801}.\\

\begin{figure}
    \begin{center}
        \includegraphics[width=0.52\textwidth]{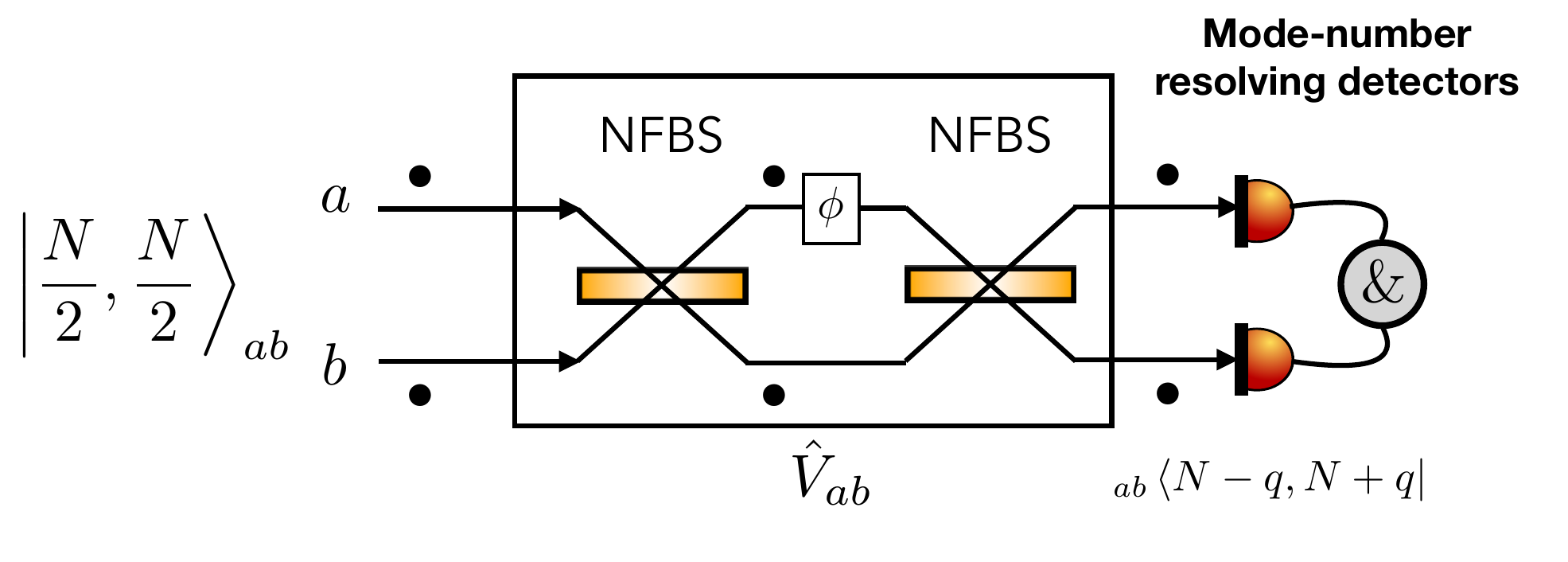}
        \caption{\label{inte2}Interferometry at the Heisenberg limit for phase estimation with a two-photon state. The protocol starts with a separable two-photon state whose spectrum is a Hermite--Gauss function of order $N$. The two photon state becomes entangled by a first non-linear frequency beam-splitter (FBS), and crosses the sample that induces a frequency shift.  A measurement in an entangled basis is then performed, that is composed of non-linear frequency beam-splitter and a mode-resolved detector.}
        \end{center}
\end{figure}

We start with the two-photon state with spectrum $\ket{N/2,N/2}_{ab}$, which corresponds to a separable two-photon state with Hermite--Gauss spectrum of order $N/2$, and is therefore mathematically analogous to a twin-Fock state \cite{holland_interferometric_1993}. Then, this two-photon state enters in the interferometer described in Fig.~\ref{inte2}, which contains a NLFBS followed by a fractional Fourier transform in the frequency domain (see Eq.~(\ref{fractional})), and finally by another NLFBS. The detection is composed of two single-photon detectors that are resolved in Hermite frequency modes. In other words, we are considering  mode-resolved single photon detectors and there are in perfect mathematical analogy with photon-number resolved detectors. \\

By using the analogy between the position-momentum quadrature and the time-frequency degree of freedom, we can proceed to the Schwinger mapping \cite{demkowicz-dobrzanski_quantum_2015} between `orbital angular momentum operators' and the time-frequency operators as follows:
\begin{align}
    \hat{J}_{x}=\frac{1}{2} (\hat{\omega}_{a}\hat{\omega}_{b}+\hat{t}_{a}\hat{t}_{b})\\
      \hat{J}_{y}=\frac{1}{2} (-\hat{\omega}_{a}\hat{t}_{b}+\hat{t}_{a}\hat{\omega}_{b})\\
    \hat{J}_{z}=\frac{1}{4}( \hat{\omega}_{a}^{2}+\hat{t}_{a}^{2}-(\hat{\omega}_{b}^{2}+\hat{t}_{b}^{2})). 
\end{align}
In the Heisenberg representation, the orbital angular momentum operators after the non-linear interferometer can be written as:
\begin{equation}
    \begin{pmatrix}
        \hat{J}_{x}\\\hat{J}_{y}\\\hat{J}_{z}
    \end{pmatrix}\mapsto\begin{pmatrix} \text{cos}(\phi) & 0 & \text{sin}(\phi) \\ 0 & 1 & 0 \\ -\text{sin}(\phi) & 0 & \text{cos}(\phi) \end{pmatrix} \begin{pmatrix}
        \hat{J}_{x}\\\hat{J}_{y}\\\hat{J}_{z}
    \end{pmatrix}.
\end{equation}
By measuring the difference between the measured mode number that is analogous of measuring the difference between the photocurrents, we access the measurement of $\hat{J}_{z}$. The phase precision can be written as:
\begin{equation}
    \Delta \phi= \frac{\Delta \hat{J}_{z}}{ \abs{d \langle \hat{J}_{z}\rangle /d\phi}},
\end{equation}
where $\langle \hat{J}_{z}\rangle=\text{cos}(\phi) \langle \hat{J}_{z}\rangle_{\text{in}}-\text{sin}(\phi) \langle \hat{J}_{x}\rangle_{\text{in}}$, and $\Delta^{2}\hat{J}_{z}=\text{cos}^{2}(\phi)\Delta^{2}\hat{J}_{z}|_{\text{in}}+\text{sin}^{2}(\phi)\Delta^{2}\hat{J}_{x}|_{\text{in}}-2\text{sin}(\phi)\text{cos}(\phi) \text{cov}(\hat{J}_{x},\hat{J}_{z})|_{\text{in}}$. For a very large mode $N$, it was shown in \cite{holland_interferometric_1993,demkowicz-dobrzanski_quantum_2015} that  the sensitivity for the phase estimation is $\Delta \phi\sim 1/N$, that scales as the inverse of the number of modes. This is an example of spectral engineering of photon pairs that is employed for increasing the resolution over the measurement of a parameter as it was done in \cite{fabre_parameter_2021}. This is a metrological scenario relevant for optically sensitive materials. Apart from the fractional Fourier transform and the mode-resolved measurement, the main challenge in this protocol lies in performing the frequency-entangling operation. \\


\section{Gaussian Boson sampling in the frequency domain of single photons}
\label{GBS}

\begin{figure*}
    \begin{center}
        \includegraphics[width=0.75\textwidth]{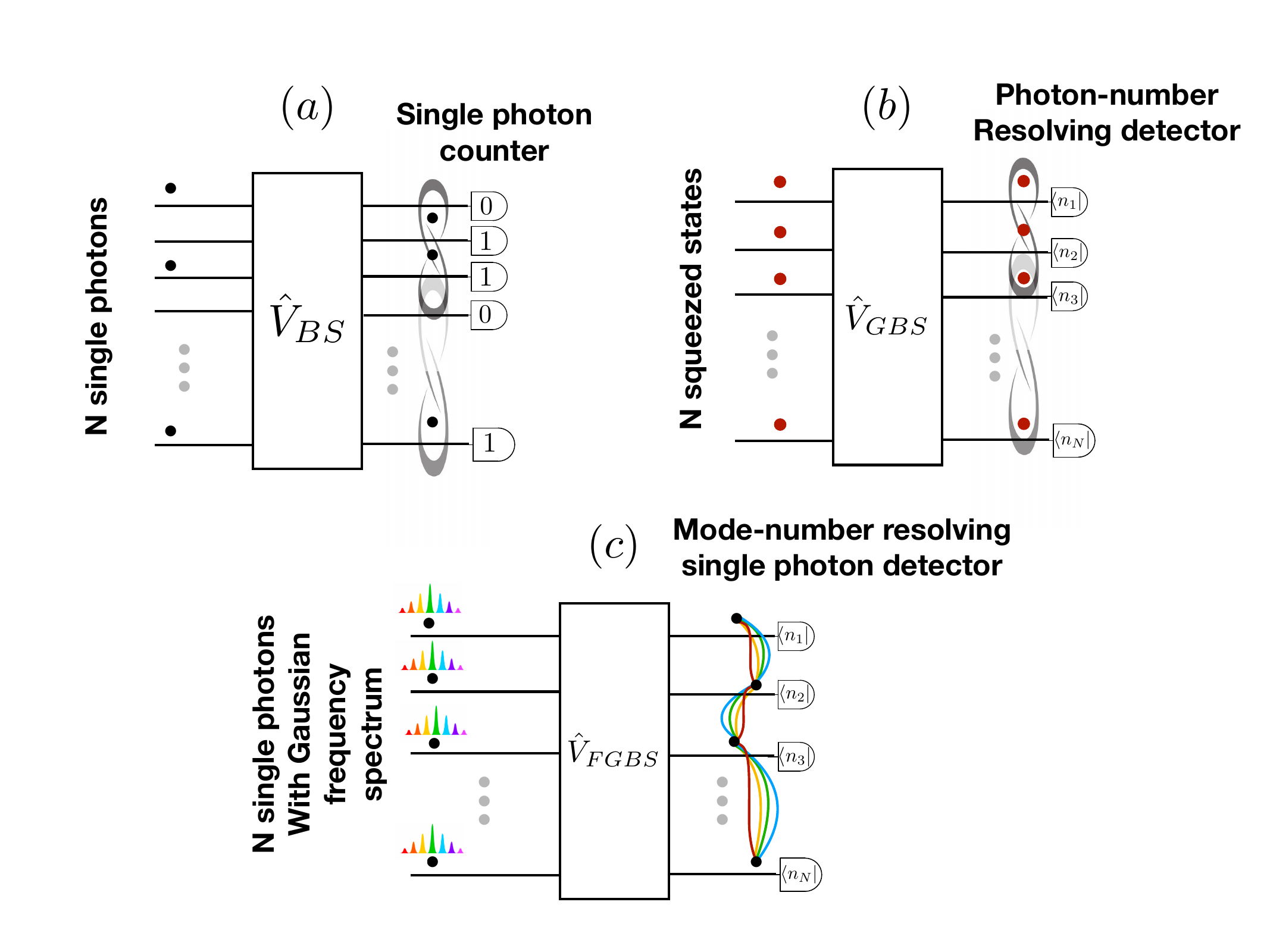}
        \caption{\label{newGBS}(a) Boson Sampling, where each black dot corresponds to a single photon occupying one single spatial mode. Each photon is indistinguishable. The output state is an entangled state of a presence/absence of single photon into $M$ spatial modes. (b) GBS, where each red dot corresponds to a single-mode squeezed state. The output state after the Gaussian operation is a multimode entangled Gaussian state that is then measured with $N$ photon number-resolving detectors. (c) FGBS experiment, where each black dot supplemented by a frequency comb indicated one single photon with a Gaussian \textit{continuous} spectral structure. The output state after the time-frequency interferometer is a multimode (spatial for instance) time-frequency entangled Gaussian state (schematized with colored lines) that is then measured with frequency mode-resolved single photon detectors. }
    \end{center}
\end{figure*}

With such equivalence between quadrature and frequency degrees of freedom of single photons, we obtain straightforwardly the equivalent of GBS in the frequency domain of single photons, which we call Frequency-based Gaussian Boson Sampling (FGBS).

\subsection{Description of the protocol}

We start from the mathematical analogue of $N$ squeezed states which are $N$ single photons with a Gaussian frequency spectrum. These states undergo time-frequency  Gaussian operations. A time-frequency Gaussian state is a state whose chronocyclic Wigner distribution is Gaussian and characterized by a covariance matrix $\Sigma$ (see Appendix~\ref{details} for more details). 
These time-frequency Gaussian operations can be achieved with linear and non-linear optics \cite{fabre_time-frequency_2022}. Measurements are performed with a set of single-photon detectors that have been frequency filtered. This allows for a mode-number resolved detection as described in the previous section. The probability distribution of detecting a total of $N$ photons (one photon at each auxiliary mode output) in the spectral configuration $\vec{n}=(n_{1},...,n_{N})$, where $n_{i}$ labels the frequency modes in the output $i$, is:
\begin{equation}
    P(\vec{n})=|\Sigma_{Q}|^{-1/2} \frac{\mathrm{haf}(A_{\vec{n}})}{\prod_{i=1}^{N} n_{i}!},
\end{equation}
where $\Sigma_{Q}=\Sigma+\mathds{I}_{2N}$ and $A=\begin{pmatrix} 0 & \mathds{I} \\ \mathds{I} & 0 \end{pmatrix}(\mathds{I}+(\Sigma+\mathds{I}/2)^{-1})$, with $A_{\vec{n}}$ the $2n\times2n$ submatrix of $A$ obtained by repeating $n_i$ times its rows and columns $i$ and $N+i$, where $n=\sum_{i}n_{i}$ is the sum of the labels of the detected modes. The function $\mathrm{haf}$ is the hafnian: for a $2n\times2n$ matrix $B$,
\begin{equation}
    \mathrm{haf}(B)=\frac{1}{n!2^{n}}\sum_{\sigma \in S_{2n}} B_{\sigma(1)\sigma(2)}\dots B_{\sigma(2n-1)\sigma(2n)},
\end{equation}
where the sum is over the permutations of $\{1,\dots,2n\}$. We give a detailed derivation in Appendix~\ref{app:FGBS}, following \cite{gbs}.

Therefore, the FGBS is a non-linear optics scheme to generate samples extracted from a mode distribution generated by a time-frequency light source at the output of a multimode interferometer. The hardness of the classical simulation of this sampling task is preserved in noisy experimental condition if the level of photon losses and temporal and frequency broadening (mathematically similar to photon losses in GBS) are limited. 

Using single photons and their continuous-variables spectral distribution in FGBS leads us to a model that "mixes" both features of Boson sampling and GBS.
In Fig.~\ref{newGBS}, we have represented the three types of bosonic samplers based on Boson sampling, GBS and FGBS. The Boson Sampling protocol involves $N$ indistinguishable single photons entering an $M$-mode interferometer (with $M>N$). Phase interference determines the exit point of each single photon from the interferometer. A configuration is identified by the detector that registers a click. GBS, as previously discussed, is mathematically equivalent to FGBS.


\subsection{Experimental considerations}

Experimentally, implementing the FGBS algorithm could involve utilizing an integrated circuit designed to take as input $n$ single photon states with a Gaussian spectrum. It is worth noting that currently, single photon sources with a frequency Gaussian spectrum are not readily available; typically, they exhibit a Lorentzian spectrum, as demonstrated in \cite{somaschi_near-optimal_2016}.

However, heralded single-photon sources offer a workaround, as they can have a Gaussian spectrum when the initial photon pair also exhibits one \cite{PhysRevA.98.043813}. This can be achieved through techniques such as engineering the spatial poling of the non-linear crystal \cite{PhysRevA.93.013801} or using an integrated AlGaAs circuit \cite{10.1063/1.5015951}.\\

Similar to the preceding sections, a fundamental element for a multimode interferometer achieving large frequency entanglement among single photons is the non-linear frequency beam-splitter, achieved experimentally in \cite{Le_Jeannic_2022} by a non-local carving operation of the joint spectral amplitude of initial separable two-photon state. We provide in Appendix \ref{appendix:FBS} the description of such a scheme.


A frequency mode-resolved detector is equivalent to a "random frequency mode sorter". A mode sorter takes an input beam containing a superposition or mixture of various optical modes and spatially (or otherwise) separates these modes so that each emerges from a different output port (see for instance \cite{fontaine_laguerre-gaussian_2019}). For example, an OAM mode sorter can direct photons carrying different orbital angular momentum values into distinct spatial channels \cite{ionicioiu_sorting_2016}. The device typically implements a tailored optical transformation—often via specially designed refractive or diffractive elements—that maps a specific mode property (like the helical phase front of OAM modes) onto a position or another easily separable degree of freedom. In some implementations, interferometric techniques or phase-correcting elements are used to enhance the separation between modes.\\

Frequency HG mode sorter can be achieved with a quantum pulse gate \cite{brecht_photon_2015} that selects the output mode. A quantum pulse gate (QPG) is a device that uses nonlinear optical processes—most notably sum-frequency generation (SFG)—to perform mode-selective operations on single-photon pulses. In essence, it is engineered to pick out (or "gate") a specific temporal or spectral mode from a multimode input and convert it to a different frequency while leaving the other modes largely unaffected. This mode-selective conversion is achieved by carefully designing the spectral and temporal properties of the pump pulse used in the nonlinear process. As the QPG is based on a non-linear process, the experimental efficiency can be low, but acts as a mode sorter of frequency HG modes. However, this is not a random mode sorter, but that could be achieved by introducing a random seed to select randomly of the input.
We note that the randomness of the seed is important, otherwise this is just a projection into "one single mode", while we should be able to measure the histogram of each mode.
The most crucial aspect of why quantum pulse gate can not be used for performing such operation is that at the output of the QPG, the spatial mode is not populated by a single photon, but by a weak coherent state. This is because the QPG is based on stimulated parametric down conversion. As the auxiliary modes (here spatial) would be populated by more than one photon by using such a QPG, is has the consequences of breaking the Heisenberg algebra checked by the time and frequency operators Eq.~(\ref{heisenberg}).\\

Therefore, to construct this spectral mode sorter, we prefer to rely on alternative techniques that avoid stimulated non)linear effect while utilizing the same optical setup used for manipulating the transverse degree of freedom. Additionally, since we operate in the frequency domain, we have to incorporated gratings.



\section{Conclusion}

In this paper, we have used the correspondence between the quadrature and the frequency degrees of freedom of single photons to design an analogue of the HOM effect in the frequency domain. In this scheme, there is a non-trivial destructive effect between different components of the frequency spectrum of two single photons, leading to a suppression of the part of the spectrum with first order Hermite--Gauss function. We have employed a conceptual representation that allows us to understand this destructive interference effect for any type of continuous variables, instead of using the standard particle-number representation. We have discussed a direct application of the correspondence to a quantum metrology protocol that can reach Heisenberg scaling for phase estimation using two photons and frequency mode resolving detection. Finally, we have discussed the equivalent of GBS in the frequency domain of single photons, established thanks to the mathematical correspondence between the position-momentum quadratures of the electromagnetic field into a single mode and the time-frequency variables of single photons. The output distribution is the hafnian of a matrix is a computationally intensive function to compute exactly, classified as a $\#$P-hard problem. Such a distribution is obtained in this manuscript with a non-linear interferometer, by contrast with the traditional Gaussian Boson sampling where the same distribution is obtained with linear optics.  

We underline the current experimental difficult to implement frequency entanglement for single photons and mode-number resolving detector. It is important to note that the proposed FGBS is not presented as a means to achieve a practical experimental advantage. Rather, it serves to illustrate that different quantum resources—despite their distinct physical implementations—can ultimately yield the same statistics.
While the physical components on which our theoretical proposal is based are challenging to implement, we expect this work to motivate the development of spectral interferometry experiments, following the example of other recent protocols \cite{PhysRevA.96.022301,folge_scheme_2024}. As a perspective, we note that our results are also valid for other types of continuous variables, such as the position and momentum of a nanomechanical resonator, or the transversal position and momentum of single photons \cite{tasca_continuous_2011}.\\


\section*{Acknowledgments}

N.\ Fabre acknowledges fruitful discussions with Olivier Pfister and Pérola Milman. U.\ Chabaud acknowledges interesting discussions with Jack G.\ Davis.

\appendix

\section{Time-frequency continuous variables of single photons}\label{details}

In this section, we recall the formalism of the time-frequency variables of single photons and refer the reader to \cite{fabre_time-frequency_2022, fabre:tel-03191301, 10.1117/12.2665062} for further details.

\subsection{Hilbert space representation}

A single photon at spatial mode $a$ with frequency $\omega$ is described by the application of the creation operator on the vacuum state: $\hat a^\dag(\omega)\ket\Omega_a=|\omega\rangle_a$. The annihilation operator is defined as follows : $\hat{a}(\omega)|\omega'\rangle=\delta(\omega-\omega')\ket\Omega_a$.  The creation and annihilation operators obey the canonical commutation relations:
\begin{equation}\label{comm}
    [\hat{a}(\omega), \hat{a}^{\dagger}(\omega')]=\delta(\omega-\omega')\hat{\mathds{I}}, 
\end{equation}
and we have also $[\hat{a}(\omega), \hat{a}(\omega')]=0$, $[\hat{a}^{\dagger}(\omega), \hat{a}^{\dagger}(\omega')]=0$. The Fourier transform of the creation (resp.\ annihilation) operator is the creation (resp.\ annihilation) operator at the arrival time $t$:
\begin{equation}
    \hat{a}(t)=\frac{1}{\sqrt{2\pi}} \int_{\mathbb{R}} d\omega \hat{a}(\omega) e^{-i\omega t}.
\end{equation}
A single photon at the arrival-time $t$ can then be defined as: $\hat{a}^{\dagger}(t)\ket{\Omega}=\ket{t}_{a}$. The spectral state of a single photon can be described by a density matrix:
\begin{equation}
    \hat{\rho}=\int_{\mathbb{R}} d\omega \int_{\mathbb{R}} d\omega' \rho(\omega,\omega') |\omega\rangle \langle\omega'|_a,
\end{equation}
which satisfies $\mathrm{Tr}(\hat{\rho})=1$. For a pure state $\hat{\rho}=|\psi\rangle \langle\psi|$, the corresponding wave function can be decomposed either in the time-of-arrival basis:
\begin{equation}\label{singlephotontime}
    |\psi\rangle_a=\int_{\mathbb{R}} dt \tilde{S}(t) |t\rangle_a,
\end{equation}
or in the frequency basis:
\begin{equation}
    |\psi\rangle_a=\int_{\mathbb{R}} d\omega S(\omega)|\omega\rangle_a,
\end{equation}
where the amplitude spectrum $S(\omega)$ is the Fourier transform of the time-of-arrival amplitude  distribution $\tilde{S}(t)$.  The relation between the spectral and temporal amplitude is valid in the case of quasi-monochromatic spectrum \cite{PhysRevA.72.032110}, i.e.\ when the central frequency of the spectral distribution is much larger than the spectral width (see Ref.\ \cite{smith_photon_2007} for the relation between the temporal and spectral distribution in the general case).  Note that the normalization of the wave function imposes $\int_{\mathbb{R}} dt  |\tilde{S}(t)|^{2}=1$.\\

A temporally (resp.\ frequency-) resolved measurement gives access to the density probability $P(t)=| \langle t | \psi\rangle |^{2}$ (resp.\ $P(\omega)=| \langle \omega | \psi\rangle |^{2}$). If the measurement is not temporally resolved, one does not get information about the time-of-arrival of the single photon. In the case of the HOM experiment, we only access the coincidences that are modified by the time delay between the two arms of the interferometer.

The time and frequency operators can be built as follows:
\begin{align}
    \hat{\omega}_{a}=\int_{\mathbb{R}} \omega d\omega  \hat{a}^{\dagger}(\omega)\hat{a}(\omega),\\
    \hat{t}_{a}=\int_{\mathbb{R}} t dt  \hat{a}^{\dagger}(t)\hat{a}(t).
\end{align}
They admit as eigenvectors the single-photon states $\hat{\omega}_{a}\ket{\omega}_{a}=\ket{\omega}_{a}$ and $\hat{t}_{a}\ket{t}_{a}=\ket{t}_{a}$. When there is only one single photon at a spatial mode $a$, the frequency and time operators obey the canonical commutation relation \cite{fabre_time-frequency_2022}:
\begin{equation}
    [\hat{\omega}_{a},\hat{t}_{a}]=\mathds{I}_{a},
\end{equation}
where $\mathds{I}_{a}$ is the identity operator in the single photon subspace. This is the basis for the analogy between the quadrature degrees of freedom and the time-frequency ones.
Hence, the space of states we consider hereafter consists of a collection of $n$ single photon states in $n$ different ancillary modes. This space will be called from now on ${\cal S}_n$, where $n$ is the number of distinguishable modes and also the number of photons (see \cite{fabre_time-frequency_2022, Descamps_2023}). It means that only cases where there is at most one photon per auxiliary mode are being considered.

A general pure state in ${\cal S}_n$  can be written as
\begin{equation}\label{State}
    |\psi\rangle=\int_{\mathds{R}} d\omega_{1}\dots\int_{\mathds{R}} d\omega_{n} F(\vec\omega) \hat{a}_{1}^{\dagger}(\omega_{1})\dots\hat{a}_{n}^{\dagger}(\omega_{n})|\Omega\rangle_a,
\end{equation}
where the spectral function $F(\vec\omega)=F(\omega_{1},\dots,\omega_{n})$ is normalized to one : $\int d\omega_{1}\dots\int d\omega_{n} | F(\omega_{1},\dots,\omega_{n})|^2 =1$. \\

We can introduce also "time-frequency coherent states" as the single photon states with the Gaussian frequency distribution:
\begin{equation}\label{timefrequencycoherent}
    \ket{\alpha=\omega_{0}/\sigma+i\sigma t_{0}}=\int d\omega e^{-(\omega-\omega_{0})^{2}/(2\sigma^{2})} e^{i\omega t_{0}} \ket{\omega},
\end{equation}
which are analogous to the (quadrature) coherent states. We can check that these states satisfy the overcompleteness relation: $\int d^{2}\alpha \ket{\alpha}\bra{\alpha}=\mathds{I}$.

\subsection{Time-frequency phase space representation}
The chronocyclic Wigner distribution has been introduced in Ref.\ \cite{fabre_generation_2020}. It satisfies the Stratonovich--Weyl axioms of phase-space distributions, and provides a detailed picture of the time-frequency properties of quantum systems. It takes the form:
\begin{equation}
    W_{\hat{\rho}}(\omega,t)=\int_{\mathds{R}} d\omega' e^{2i\omega' t} \langle \omega-\omega'| \hat{\rho}|\omega+\omega'\rangle.
\end{equation}
The integration is taken over the real line as soon as the frequency distribution obeys the narrow band condition.  The chronocyclic Wigner distribution can be expressed as the average value of the displaced parity operator:
\begin{equation}
    W_{\hat{\rho}}(\omega,t)=\text{Tr}(\hat{\rho} \hat{\Pi}(\omega,t)) \ , \ \hat{\Pi}(\omega,t)=\hat{D}(\omega,t)\hat{\Pi} \hat{D}^{\dagger}(\omega,t),
\end{equation}
where the parity operator is $\hat{\Pi}=\int_{\mathds{R}} d\omega |\omega+\omega_{p} \rangle \langle \omega_{p}-\omega |$, where $\omega_{p}$ is the central frequency corresponding to the origin of the chronocyclic phase space. The displacement operators $\hat{D}(\omega,t)=e^{-i\omega t/2}\int d\omega' e^{i\omega' t} |\omega+\omega'\rangle \langle \omega' |$ obey Weyl commutation relations, where the non-commutativity comes from the non-commutation of bosonic operators (see Eq.~(\ref{comm})). There is a one-to-one correspondence between the chronocyclic Wigner distribution and the matrix element of the density matrix, called the (chronocyclic) Weyl transform:
\begin{equation}
    \rho(\omega,\omega')=\int_{\mathds{R}} dt e^{i(\omega-\omega')t} W_{\hat{\rho}}\left(\frac{\omega+\omega'}{2},t\right).
\end{equation}
The chronocyclic Wigner distribution  is normalized to one, $\iint d\omega dt  W_{\hat{\rho}}(\omega,t)=1$, which is a consequence of the normalization of the density matrix. The marginals of the chronocyclic Wigner distribution correspond to the spectral and time-of-arrival distributions which are directly measurable experimentally:
\begin{equation}
    \int_{\mathds{R}} d\omega W_{\hat{\rho}}(\omega,t)= \langle t | \hat{\rho} | t \rangle  \ , \ 
    \int_{\mathds{R}} dt W_{\hat{\rho}}(\omega,t)= \langle \omega | \hat{\rho} | \omega \rangle .
\end{equation}
Moments of the distribution can be expressed either with a quantum average or a statistical one
\begin{align}\label{variancephasematc}
\langle \omega^{\alpha} t^{\beta} \rangle =\langle \hat{\omega}^{\alpha}\hat{t}^{\beta}\rangle= \int_{\mathds{R}}  \int_{\mathds{R}} d\omega dt \omega^{\alpha} t^{\beta} W_{\hat{\rho}}(\omega,t).
\end{align}
The chronocyclic Wigner distribution can also be defined for $n$ excitations of the electromagnetic field, 
\begin{multline}
    W_{\hat{\rho}}(\omega_{1},t_{1}; \omega_{2},t_{2};\dots; \omega_{n},t_{n})=\\
    \int_{\mathds{R}} d\omega'_{1}\dots\int_{\mathds{R}} d\omega'_{n} e^{2i(\omega_{1}' t_{1}+\dots+\omega_{n}'t_{n})} \\
    \times\langle \omega_{1}-\omega_{1}',\dots,\omega_{n}-\omega_{n}'|\hat{\rho}|\omega_{1}+\omega_{1}',\dots,\omega_{n}+\omega_{n}'\rangle.
\end{multline}
However, note that each spatial mode is occupied by one single photon. By considering a single photon pure state $\hat{\rho}=|\psi\rangle \langle \psi|$ and $|\psi\rangle=\int S(\omega) d\omega |\omega\rangle$, the chronocyclic Wigner distribution can be written as:
 \begin{equation}
    W_{\hat{\rho}}(\omega,t)=\int_{\mathds{R}} d\omega' e^{2i\omega' t}  S(\omega-\omega')S^{*}(\omega+\omega').
 \end{equation}
We can straightforwardly introduce the time-frequency Husimi distribution and the P-distribution of spectral single photon states using the time-frequency coherent state (see Eq.(\ref{timefrequencycoherent}):
\begin{align}
        Q_{\rho}(\alpha)=\bra{\alpha} \hat{\rho} \ket{\alpha}\\
        \hat{\rho}=\iint d^{2}\alpha P_{\hat{\rho}}(\alpha) \ket{\alpha}\bra{\alpha}.
\end{align}
We use these definitions in Appendix~\ref{app:FGBS} to derive probability expressions for Frequency-based Gaussian Boson Sampling.

\subsection{Time-frequency Gaussian states}

Now, we describe time-frequency Gaussian states \cite{fabre:tel-03191301}, following the quadrature ones \cite{braunstein_quantum_2005}. We introduce the column vector  $X=(\omega_{1},t_{1},\dots,\omega_{n}, t_{n})^T$, where $1,2,\dots,n$ denote $n$ spatial (or other discrete auxiliary) modes. We consider that each spatial region is not overlapping, so that $[\hat{\omega}_{i},\hat{t}_{j}]=i\hat{\Omega}_{ij}$ with $\hat{\Omega}=\begin{pmatrix} \mathds{I}& 0\\ 0 &-\mathds{I} \end{pmatrix}$. The displacement vector is $\overline{X}_{i}=\text{Tr}(\hat{\rho} \hat{X}_{i})$. Assuming a pure state, we obtain that the displaced vector is the average value of the Gaussian distribution. We can define the covariance matrix as:
\begin{equation}
    \Sigma_{ij}=\text{Tr}(\hat{\rho} \{ \Delta \hat{X}_{i}, \Delta \hat{X}_{j}^{\dagger} \} ),
\end{equation}
where $\Delta \hat{X}_{i}=\hat{X}_{i}^{2}-\langle \hat{X}_{i} \rangle^{2}$. A Gaussian state is a state only characterized by its first two moments, and whose Wigner distribution is Gaussian. 
The chronocyclic Wigner distribution of a time-frequency Gaussian state $\hat{\rho}$ can be written as:
\begin{equation}
    W_{\hat{\rho}}(\omega,t)=\text{exp}(-(X-\overline{X})^{T} \Sigma^{-1} (X-\overline{X})).
\end{equation}
Time-frequency Gaussian transformations are unitary operators generated by Hamiltonians which are at most quadratic in $X$. Equivalently, they can be described by symplectic matrices, and we give a few examples in what follows.

Rotation operations can be cast under the form: $\hat{R}(\phi)=e^{i\phi(\hat \omega^2 + \hat t^2)}$ which is a fractional Fourier transform. Similarly, the action of the squeezing operator modifies the frequency bandwidth of a single photon state.

The non-linear frequency beam-splitter is also a time-frequency Gaussian operation, as it is quadratic in the time and frequency operators, that can be cast as follows:
\begin{equation}\label{lal}
    \hat{U}=\text{exp}(i\pi/4 (\hat{\omega}_{a}\otimes \hat{t}_{b}-\hat{t}_{a}\otimes \hat{\omega}_{b})).
\end{equation}

Finally, the chronocyclic Husimi distribution of a time-frequency Gaussian state is Gaussian and can be cast as:
\begin{equation}
    Q_{\hat{\rho}}(\omega,t)=\text{exp}(-(X-\overline{X})^{T} \Sigma (X-\overline{X})).
\end{equation}

\section{The non-linear frequency beam-splitter}\label{appendix:FBS}

Let us begin by providing a precise formulation of the NLFBS. The unitary operator for this beam-splitter is expressed as:

\begin{equation}\label{untiary}
\hat{U} = \iint d\omega \, d\omega' \, \hat{a}^{\dagger}\left(\frac{\omega + \omega'}{\sqrt{2}}\right)\hat{b}^{\dagger}\left(\frac{\omega - \omega'}{\sqrt{2}}\right) \hat{a}(\omega)\hat{b}(\omega'), 
\end{equation}
where \( \hat{a} \) and \( \hat{b} \) represent two auxiliary modes, each occupied by precisely one single photon. The Eq.~(\ref{untiary}) is equivalent to Eq.~(\ref{lal}) when each mode $a,b$ are populated with one single photon (see also \cite{fabre_time-frequency_2022}). This operation effectively performs a rotation of the joint spectral amplitude (JSA) of photon pairs by \(\pi/4\) radians. 

To illustrate this, consider a Gaussian separable JSA state as the input to the non-linear frequency beam-splitter:

\begin{equation}
\ket{\psi} = \iint d\omega_{s} \, d\omega_{i} \, \exp\left(-\frac{\omega_{s}^2}{2\sigma^2}\right)\exp\left(-\frac{\omega_{i}^2}{2\sigma'^2}\right) \ket{\omega_{s}, \omega_{i}}_{ab},
\end{equation}
where \( \sigma \) and \( \sigma' \) denote the Gaussian widths of the spectral distributions of the photons. After applying the non-linear frequency beam-splitter, the two-photon state evolves to:

\begin{equation}
\begin{aligned}
\hat{U}\ket{\psi} = \iint d\omega_{s} \, d\omega_{i} \, &\exp\left(-\frac{\omega_{s}^2}{2\sigma^2}\right)\exp\left(-\frac{\omega_{i}^2}{2\sigma'^2}\right)\\&\times\ket{\frac{\omega_{s} + \omega_{i}}{\sqrt{2}}, \frac{\omega_{s} - \omega_{i}}{\sqrt{2}}}_{ab}.
\end{aligned}
\end{equation}
By performing a change of variable, this state can be rewritten as:

\begin{equation}
\begin{aligned}
\hat{U}\ket{\psi} = \iint d\omega_{s} \, d\omega_{i} \, &\exp\left(-\frac{\left(\frac{\omega_{s} + \omega_{i}}{\sqrt{2}}\right)^2}{2\sigma^2}\right) \exp\left(-\frac{\left(\frac{\omega_{s} - \omega_{i}}{\sqrt{2}}\right)^2}{2\sigma'^2}\right)\\&\times\ket{\omega_{s}, \omega_{i}}_{ab}.
\end{aligned}
\end{equation}

Here, the joint spectral function is now dependent on the collective variables \(\omega_{\pm} = \frac{\omega_{s} \pm \omega_{i}}{\sqrt{2}}\), rather than the local variables, indicating the emergence of frequency entanglement. This transformation leads to an entangled wavefunction in the frequency domain (see the right hand side of Fig.~\ref{fbsss}).\\

We now mention two potential methods for implementing a non-linear frequency beam-splitter:

(1) Joint spectral amplitude carving (or non-local filtering):
   This method has been experimentally demonstrated using the linear coupling of a quantum dot embedded within a waveguide \cite{Le_Jeannic_2022}  and proposed theoretically with quadratic coupling \cite{PRXQuantum.4.030326}. The process involves modifying the initial circular JSA of two separable single photons to an entangled state through non-local filtering induced by light-matter interactions. The non-local filtering comes from the scattering effect of the two single photons into the quantum dot. 

(2) Performing the true rotation of the JSA. This has not been implemented experimentally in photonics platforms yet. However, we have found similar operations for superconducting circuits \cite{chang_observation_2020,eriksson_universal_2023} and trapped ions \cite{bazavan_squeezing_2024} which have achieved tri- and quad-squeezing operations that are cubic or quartic in canonical bosonic operators as the non-linear frequency beam-splitter. Being able to design such a device in the photonic setting using in particular quantum dots and waveguides is current work in progress.\\

We note that the number of Schmidt modes in the two methods is the same, as it depends on the ellipticity of the JSA (see Fig.~\ref{fbsss}).

\section{Output probability distribution for Frequency-based Gaussian Boson Sampling}
\label{app:FGBS}

In this section, we give a derivation for the output probability distribution of a FGBS device. Once the mathematical analogy between the time-frequency and the position-momentum quadratures is established, we can conceive the same interferometer for the two type of degrees of freedom that would lead to the same output distribution, and thus the derivation is identical to that in \cite{gbs}.

The $N$-mode time-frequency Gaussian state is then measured with the $N$ mode-number sensitive detector, described by the projector $\ket{n_{1}}\bra{n_{1}}...\ket{n_{N}}\bra{n_{N}}$. The probability of measuring one single photon each mode $n_{1}...n_{N}$ is therefore given by $P(n_{1},...,n_{N})=\abs{\bra{n_{1},...,n_{N}}\ket{\psi}}^{2}$ when the state is pure: 
\begin{equation}
P(n_{1},...,n_{N})= \pi^{N} \int d^{2N} \alpha Q_{\hat{\rho}}(\vec{\alpha}) P_{\vec{n}}(\alpha),
\end{equation}
where $d^{2N} \alpha=\prod_{j=1}^{M} d\alpha_{j}d\alpha^{*}_{j}$. After integrations by parts, it leads to
\begin{equation}
P(n_{1},...,n_{N})=\frac{1}{n!\sqrt{|\sigma_Q|}} \prod_{j=1}^{M} \left(\frac{\partial^2}{\partial\alpha_j\partial\alpha_j^*}\right)^{n_j} e^{\frac{1}{2}\alpha_v^{\dagger}A\alpha_v}\bigg|_{\alpha_v=0},
\end{equation}
where $\sigma_Q = \sigma + \mathds{I}_{2M}/2, \quad \alpha_v = [\alpha_1,\dots,\alpha_M,\alpha^{*}_1,\dots,\alpha^{*}_M]$
\begin{equation}
A = \begin{pmatrix}
0 & \mathds{I}_M\\
\mathds{I}_M & 0
\end{pmatrix} \begin{bmatrix}
\mathds{I}_{2M} - \Sigma_Q^{-1}
\end{bmatrix}.
\end{equation}
The matrix $\sigma$ contains only the modes that are measured, while the unobserved modes are traced over to obtain a reduced covariance matrix. Restricting to an even number of photons (consequence of the Wick's theorem), we obtain
\begin{equation}
P(n_{1},...,n_{n})=
\left\lbrace
\begin{array}{ccc}
\abs{\text{haf}(A)}^{2} & \mbox{if} & \Omega=\mbox{even} \\
0 & \mbox{if} & \Omega=\mbox{odd},  \\
\end{array}\right.
\end{equation}
where $\Omega:=n_1+\dots+n_n$ and where the hafnian of the symmetric matrix $A$ of size $n=2m$ is defined as:
\begin{equation}
\text{haf}(A)=\frac{1}{m!2^{m}}\sum_{\sigma \in S_{n}} A_{\sigma(1)\sigma(2)}...A_{\sigma(n-1)\sigma(n)},
\end{equation}
where the sum is over the permutations of $\{1,\dots,n\}$. Therefore, the hafnian, denoted as \(\text{haf}(A)\), is defined analogously to the permanent of a matrix but is specific to symmetric matrices and involves a sum over perfect matchings. Mathematically, if \( A = (a_{ij}) \) is a symmetric matrix, the hafnian of \( A \) is defined as:
\begin{equation}
 \text{haf}(A) = \sum_{\text{matchings}} \prod_{(i,j) \in \text{matching}} a_{ij},
\end{equation}
where the sum is over all possible perfect matchings of the set \(\{1, 2, \ldots, n\}\). A perfect matching is a way of pairing up the elements so that each element is paired exactly once with another element. To provide a concrete example, consider a \( 4 \times 4 \) symmetric matrix \( A \):

\begin{equation}
    A = \begin{pmatrix}
a_{11} & a_{12} & a_{13} & a_{14} \\
a_{12} & a_{22} & a_{23} & a_{24} \\
a_{13} & a_{23} & a_{33} & a_{34} \\
a_{14} & a_{24} & a_{34} & a_{44}
\end{pmatrix}.
\end{equation}
The hafnian of \( A \) would be computed by summing over the products of the matrix elements corresponding to all perfect matchings:

\begin{equation}
\text{haf}(A) = a_{12}a_{34} + a_{13}a_{24} + a_{14}a_{23}.
\end{equation}
This sum considers all possible ways to pair up the indices \( \{1, 2, 3, 4\} \) such that each index is used exactly once.





\bibliographystyle{apsrev4-2}
\bibliography{GBS1}


\end{document}